# A geometric method to determine the direction of electric field due to a uniformly charged line segment

Fulin Zuo

Department of Physics

University of Miami

Coral Gables, FL 33124

## Abstract

A geometrical approach to calculate the electric field due to a uniformly charged rod is presented. The result is surprisingly simple and elegant. Using pure geometrical quantities like length and angle, the direction of the electric field in this charge configuration can be graphically defined. Understanding and application of this result can lead to a deeper appreciation of symmetry in a seemingly un-symmetric system.

## Introduction:

The electromagnetism part of the college physics is widely perceived as the most difficult part of the introductory physics course where calculus and laws of physics, in this case, Coulomb's law, are combined to find quantities such as the electric field or electric potential for extended charge distributions. For most students, it is their first time to actually apply or re-learn calculus in the world of physics. To accomplish this goal, it is a gold standard in nearly every calculus based physics textbook to start with the electric field calculation for a uniformly charged line segment [1-6]. The results can be simplified when the line is infinitely long, which can then be compared to a simple algebraic derivation using Gauss's law.

Applications of Gauss's law require high symmetry of the electric field. Spherical charge distribution is the only system where full spherical symmetry is preserved for any sized system. Cylindrical or planar symmetry requires the system to be infinitely large such as the infinitely long wire or infinitely large sheet to make use of Gauss's law. For a finite uniformly charged line segment, symmetry is only defined on the axis bisecting the line. For points away from the axis, direction of the electric field is no longer well defined intuitively. While it is straight forward to calculate the components, the results often look complicated.

This naturally leads to a question: are there hidden symmetries in the problem?

## Results:

The problem here is to find the electric field everywhere due to a thin rod of length L with a uniform charge density $\lambda$. For generality, the rod is placed on the x-axis from x=a to x=b with b=a+L. The calculation is normally done for two cases, one for the field along the axis and the other away from the axis. The calculation of electric field on the x-axis is fairly straight forward, as the direction is the same for every infinitesimal contribution dE from a dq on the charged rod with $E(x') = \int_a^b \frac{k\,dq}{r^2} = \int_a^b \frac{k\lambda\,dx}{(x'-x)^2}$.

The calculation of electric field at a point P(y) on the y-axis is more involved because the finite components in both x and y directions.

There are usually two approaches to find the net electric field. Mostly, it is done by integrating $dE_x$ and $dE_y$ components in terms of variable x along the x-axis.

$$E_x = -\int_a^b \frac{kdq}{r^2}\sin\theta = -\int_a^b \frac{k\lambda dx}{r^2}\frac{x}{r} = -\int_a^b \frac{k\lambda x dx}{(x^2+y^2)^{\frac{3}{2}}} = k\lambda[\frac{1}{\sqrt{a^2+y^2}} - \frac{1}{\sqrt{b^2+y^2}}] \quad (1)$$

$$E_y = \int_a^b \frac{kdq}{r^2}\cos\theta = \int_a^b \frac{k\lambda dx}{r^2}\frac{y}{r} = \frac{k\lambda}{y}[\frac{b}{\sqrt{b^2+y^2}} - \frac{a}{\sqrt{a^2+y^2}}], \quad (2)$$

here it involves a slightly more complicated integral

$$\int \frac{dx}{(x^2+y^2)^{\frac{3}{2}}} = \frac{1}{y^2}\frac{x}{\sqrt{x^2+y^2}}$$

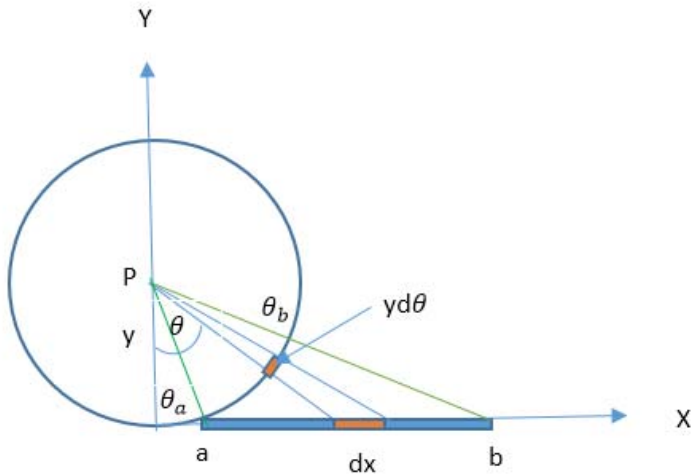

Figure 1 Calculation of electric field at point P. The infinitesimal contribution from dx is mapped to an infinitesimal arc segment $yd\theta$.

An alternative method to get around the integral above is through a change of variable from x to the angle $\theta$. By using the relationship $x = y\tan\theta$, $r = \frac{y}{\cos\theta}$, $dx = \frac{y}{(\cos\theta)^2}d\theta$, the components can be obtained as follows,

$$E_x = -\int_a^b \frac{kdq}{r^2}\sin\theta = -\int_{\theta_a}^{\theta_b} \frac{k\lambda}{y}\sin\theta\, d\theta = \frac{k\lambda}{y}[\cos\theta_b - \cos\theta_a] \quad (3)$$

$$E_y = \int_a^b \frac{kdq}{r^2}\cos\theta = \int_{\theta_a}^{\theta_b} \frac{k\lambda}{y}\cos\theta\, d\theta = \frac{k\lambda}{y}[\sin\theta_b - \sin\theta_a] \quad (4)$$

Clearly, the integral becomes much simpler in terms of the angle $\theta$. The results are of course consistent with eq. (1) and eq. (2), here $\sin\theta_b = \frac{b}{\sqrt{b^2+y^2}}$ and $\sin\theta_a = \frac{a}{\sqrt{a^2+y^2}}$.

Normally this is how this example in most college physics textbooks is done. Limiting cases are then discussed when the rod is infinitely long (L>>y) or the point of observation is far greater compared to the size of rod (y>>L). For intermediate cases, professors and students alike are generally satisfied with knowing how to solve the problem but believing the answer is too complicated to contemplate or to make sense of.

Here it is shown that the answers above for the finite rod are in fact very simple and it has a simple geometrical meaning to it. The infinitesimal contribution dE at a point P on the y-axis from $\lambda dx$ on the x-axis is given by

$$dE = \frac{kdq}{r^2} = \frac{k\lambda dx}{r^2} = \frac{k\lambda \frac{yd\theta}{(\cos\theta)^2}}{(\frac{y}{\cos\theta})^2} = \frac{k\lambda y d\theta}{y^2} \quad (5)$$

The last step in eq. (5) corresponds to the infinitesimal contribution from an arc segment of radius y and arc length $yd\theta$ with the same line charge density of $\lambda$. In another word, the electric field contribution from the charge on the x-axis can be mapped to the contribution by the charges on the circular segment of radius y, as shown in Figure 1. For a circular arc, the symmetry axis is well defined. Thus, the total electric field due to a circular segment is along the direction bisecting the arc. If the lines connecting the ends are defined by $\theta_a$ and $\theta_b$, the arc is then defined by angular spread of $\theta_b - \theta_a$, the bisecting line will be pointing at $\theta_a + \frac{1}{2}(\theta_b - \theta_a) = \frac{1}{2}(\theta_b + \theta_a)$.

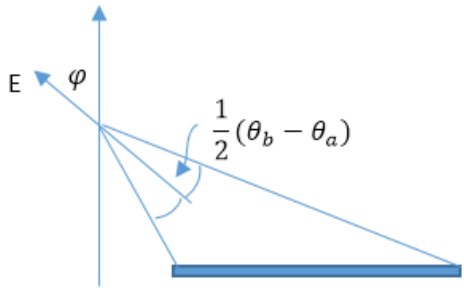

*Figure 2 E field points in the same direction as the line bisecting the angle defined by the point and the ends of the rod.*

The geometric argument is certainly consistent with the direction of total electric field calculated above (eq. (3) and eq. (4)) defined by $\cot\varphi = \frac{E_y}{-E_x} = -\frac{\sin\theta_b - \sin\theta_a}{\cos\theta_b - \cos\theta_a} = \cot\frac{1}{2}(\theta_b + \theta_a)$, as shown in Figure 2.

The magnitude of the total electric field can be calculated for the simple arc to be given by $E = \frac{2k\lambda}{y}\sin\frac{1}{2}(\theta_b - \theta_a)$ or directly from eq. (3) and eq. (4).

The simplicity of the transformation enables one to pinpoint the direction and magnitude of the electric field due to any uniformly charge rod at any point in space, using geometrically defined quantities like radius y (vertical distance from the point to the line) and the angle defined between the lines connecting the point to the ends of the rod.

One interesting example of this transformation is the calculation of electric field on the y-axis due to a semi-infinite uniformly charged wire, as shown in Figure 3. In this case, $\theta_b = \frac{\pi}{2}, \theta_a = 0$, the electric field always points in the direction of $\frac{\pi}{4}$ with a magnitude of $\frac{\sqrt{2}k\lambda}{y}$. The results look somewhat counter intuitive, as one might expect the field to point more toward the negative x-axis as y approaches zero. However, in terms of the mapped charge distribution, it corresponds to a quarter circle regardless of the radius y, so the field always points in the same direction.

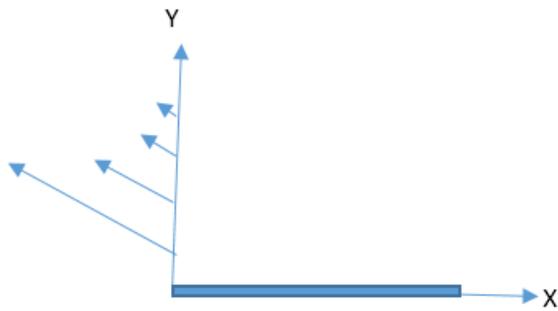

*Figure 3 Electric Field on the y-axis due to a semi-infinite wire.*

For an infinite long wire, the transformation becomes a half circle of radius y with $E=\frac{2k\lambda}{y}$, which corresponds to the result obtained easily from Gauss's law.

The direction of the electric field can also be derived from calculating the electric potential first. The gradient of the electric potential defines the direction of electric field. It is known that the equipotential surface of a charged rod corresponds to an ellipsoid. However, the mathematical transformation is rather complicated and well beyond the level of introductory physics.

The mapping transformation can be applied directly to the calculation of gravitational force due to a rod of uniform mass density on a point mass. It would be of great interest if similar projection can be found in other systems or in higher dimensions.

## Conclusion:

The contribution to electric field from an infinitesimal line segment dx can be mapped to infinitesimal arc segment of a fixed radius and this leads to a mapping of finite line segment to a finite circular arc. The symmetry axis of the arc is easily defined, thus pinpointing the direction and the magnitude of the total electric field.

It is hard to believe or even unimaginable that this has not already been found or published somewhere in the old manuscripts or hand written notes. Yet the fact that such a simple and elegant transformation is not preserved, if known, or discussed in any contemporary physics textbooks is simply astonishing. Students at the calculus level deserve and are prepared to appreciate the beauty of the transformation.

The treatment of this problem by transforming a seemingly complicated and non-symmetric charge distribution to a symmetric circular segment is interesting by itself. Discussion of this approach will enlighten and encourage science and engineering oriented students to pursue simplicity and symmetry in this complex world.

## Acknowledgement:

I want to thank many of my colleagues in the physics department and Dr. Xuewen Wang for many useful discussions. Specifically I want to thank Dr. Thomas Curtright for pointing out the ellipsoidal equipotential surface and other stimulating conversations.

## References

1. Young, Hugh D., Freedman, Roger A., Ford, A. Lewis, *University Physics with modern physics,* 13$^{th}$ Ed. (Addison-Wesley, 2011), p. 706-707.

2. David Halliday, Robert Resnick, Jearl Walker, *Fundamentals of Physics Extended*, 10$^{th}$ Ed. (Wiley, 2014), p. 643.

3. Douglas C. Giancoli*, Physics for Scientists and Engineers*, 4$^{th}$ Ed. (Pearson, 2007), p. 573.

4. Randall D. Knight, *Physics for Scientists and Engineers: A Strategic Approach*, 3rd ed. (Pearson, 2008), p. 827-828.

5. Paul Tipler, *Physics for scientists and engineers*, 3rd ed. (Worth Publishers, 1995), p. 625-628.

6. Paul M. Fishbane, Stephen Gasiorowicz, Stephen T. Thornton, *Physics for scientists and engineers* (Prentice Hall, 1993), p.682-683.